\documentclass[12pt]{iopart}
\usepackage{graphicx}
\begin{document}
\title{Triplet correlations in two-dimensional colloidal model liquids}
\author{Carsten Ru\ss{}, Klaus Zahn \ and Hans-Hennig von Gr\"unberg}
\address{Fachbereich Physik, Universit\"at Konstanz, 78457 Konstanz, 
Germany}
\begin{abstract}
  Three-body distribution functions in classical fluids have been
  theoretically investigated many times, but have never been measured
  directly. We present experimental three-point correlation functions
  that are computed from particle configurations measured by means of
  video-microscopy in two types of quasi-two-dimensional colloidal
  model fluids: a system of charged colloidal particles and a system
  of paramagnetic colloids.  In the first system the particles
  interact via a Yukawa potential, in the second via a potential
  $\Gamma/r^{3}$. Varying the particles density in the charged system,
  or, the interaction strength $\Gamma$ in the magnetic system, one
  can systematically explore how triplet correlations behave if the
  coupling between the particles changes. We find for both systems
  very similar results: on increasing the coupling between the
  particles one observes the gradual formation of a crystal-like local
  order due to triplet correlations, even though the system is still
  deep inside the fluid phase. These are mainly packing effects as is
  evident from the close resemblance between the results for the two
  systems having completely different pair-interaction potentials. To
  demonstrate that triplet correlations are significant not only
  locally, but also when integrated over the whole volume we consider
  the Born-Green equation and show that in a strongly interacting
  system this equation can be satisfied only with the full triplet
  correlation function but not with three-body distribution functions
  obtained from superposing pair-correlations (Kirkwood superposition
  approximation).
\end{abstract}
\pacs{61.20.-p, 61.20.Ne, 82.70.Dd}
\submitto{\JPCM}
\section{Introduction}

The static structure of a simple fluid is commonly described in terms
of the $m$-body distribution functions $g^{(m)}$, measuring the
probability densities of finding two, three, and more particles at
specified positions in space. The present paper is concerned with
three-body distribution functions, as obtained from measured
configurations in two-dimensional colloidal model fluids.

There are many examples in statistical mechanics where triplet
correlations in classical fluids are important. They play an essential
role, for example, in earlier theories of the pair correlation
function of a fluid, such as the Born-Green-Yvon integral equation
theory, which explicitly shows the relationsship between pair and
triplet correlation functions \cite{kirkwood35,born46,yvon35}. Triplet
correlations are furthermore used in perturbation theories for static
fluid properties \cite{stell74,madden78,gray78}, in theories of
transport properties \cite{scherwinski90}, but also to describe
solvent reorganization processes around solutes \cite{lazaridis00} or
systems under shear-flow \cite{dhont7599,dhont7710}. And, finally, one
is again led to triplet correlation functions if one is interested in
the temperature or density derivative of the pair-correlation function
$g^{(2)}(r) \equiv g(r)$ \cite{schofield66,egelbuch}.

For these and other reasons, it is not surprising to see how often
triplet correlations have been theoretically investigated, in the
overwhelming number of cases, by computer simulation studies.  Of
central importance in these studies is Kirkwood's idea
\cite{kirkwood35} to approximate the triplet distribution function by
three different pair distribution functions (the Kirkwood
superposition approximation (KSA) explained in more detail below).
The reliability and quality of the KSA has been tested extensively for
pure hard-sphere systems \cite{alder64,rice65,muller93}, for
Lennard-Jones (LJ) fluids
\cite{rahman64,raveche72,gupta82,krumhansl72a,krumhansl72b,wang72,sane82,raveche-rev,mcneil83},
for liquid sodium \cite{tanaka75}, but also for Yukawa fluids
\cite{medina98}, with the general result that all qualitative features
of $g^{(3)}$ are usually well described by the KSA. However,
quantitative differences at small distances and high densities can be
appreciable, see, for instance, \cite{mcneil83}. Mc Neil et al.
\cite{mcneil83} compared their simulations also with analytical
theories going beyond the KSA. More recent simulations aiming at an
understanding of higher-order correlations are performed in
2-component asymmetric electrolyte \cite{linse1991}, in an aqueous 1:1
electrolyte \cite{hummer92}, and in strong 3:3 and 1:3 electrolytes
\cite{jorge2002}. Finally, one has to mention theoretical work done
using integral equation theory \cite{attard91,fushiki91,raveche72},
see \cite{raveche-rev} for a review.

All these theoretical efforts contrast with the situation on the
experimental side where there are only very few papers dealing with
three-body correlations.  One indirect way to obtain information on
$g^{(3)}$ is via the isothermal pressure derivative of the fluid structure
factor $\partial S(q)/\partial P$ which is related to the triplet
distribution function \cite{schofield66}. This idea has been exploited
in rare-gas fluids mainly by Egelstaff and coworkers in a series of
experimental papers: in Krypton \cite{egel73}, Argon \cite{egel69},
Helium \cite{montfrooij91} and Neon \cite{egel273}. But, to our
knowledge, there is no direct measurement of $g^{(3)}$. It is clear
where the problems come from: x-ray or neutron diffraction
data provide information just on the fluid structure factor, i.e.,
essentially on the pair-correlation function $g(r)$.
A measurement of $g^{(3)}$, however, requires the knowledge of the
positions of three particles at the same time which is technically
very demanding to obtain in 3D samples (see however
\cite{dietrich}). 

Here lies the advantage of video-microscopy, a modern experimental
technique applied to colloidal systems to directly measure all
particles' positions at all times. Essentially, one follows the
phase-space trajectory of the system as it evolves in time, and has
thus the same amount of information as one obtains in a simulation.
It is rather straightforward to calculate higher-order correlation
functions from the measured configurations.  In \cite{russPRL}, we
have recently realized this idea and calculated three-body correlation
functions from configurations obtained by video-microscopy from a
two-dimensional system of magnetic colloids. This is probably the
first direct measurement of a three-body correlation function. In the
present paper now we give details not mentioned in \cite{russPRL},
outline a few technical points and present calculations of $g^{(3)}$
also for 2D Yukawa systems. 

\section{Theoretical background}

\subsection{Distribution functions}
The radial pair and three-body distribution functions $g(r)$ and
$g^{(3)}(\mathbf{r}_1, \mathbf{r}_2, \mathbf{r}_3)$ are used to
describe the local structure of a fluid and can best be defined by means of the
molecular distribution functions $n^{(m)}(\mathbf{r}^m)$
\cite{egelbuch}. The lowest order member of this class of functions,
$n^{(1)}(\mathbf{r}_1) dr_1$, is the probability of finding a
particle in the volume $dr_1$ at $\mathbf{r}_1$. In an homogeneous
medium it is equal to the density $\rho$.  $n^{(2)}(\mathbf{r}_1,
\mathbf{r}_2) dr_1 dr_2$ is the joint probability of finding one
particle in the volume $dr_1$ at $\mathbf{r}_1$ {\it and} a second in
the volume $dr_2$ at $\mathbf{r}_2$ and so on for higher orders. 
Dividing out the asymptotic dependency we obtain the set of $m$-body
distribution functions $g^{(m)}(\mathbf{r}^m)$:
\begin{equation}
g^{(m)}(\mathbf{r}^m) = \rho ^{-m} n^{(m)}(\mathbf{r}^m).
\end{equation}
The first non-trivial radial distribution function is 
$g(\mathbf{r}_1, \mathbf{r}_2)$ which in a homogeneous, isotropic system
depends only on the particle separation $r=|\mathbf{r}_2-
\mathbf{r}_1|$. Next in order is $g^{(3)} (\mathbf{r}_1, \mathbf{r}_2,
\mathbf{r}_3)$ describing the probability of finding triplets of
particles. Configurations of three particles are uniquely
characterized by three independent parameters (in a homogeneous,
isotropic medium), which can be chosen to be the distances between the
particles $r=|\mathbf{r}_{2}-\mathbf{r}_1|$,
$s=|\mathbf{r}_{3}-\mathbf{r}_1|$ and $t=|\mathbf{r}_3-\mathbf{r}_2|$.
Therefore, $g^{(3)} =g^{(3)} (r,s,t)$.

The potential of mean force is defined by
\begin{equation}
\label{eq:potMeanF}
\beta w^{(m)} = - \ln g^{(m)}
\end{equation}
with $\beta =1 / kT$ for the inverse thermal energy. The potentials of
mean force have to be distinguished from the direct potentials; so,
$w^{(2)}(r)$ should not be confused with the direct pair-potential
$u(r)$ in the system. The difference between both quantities can best
be understood by means of the Born-Green equation
\cite{born46,egelbuch},
\begin{equation}
  \label{eq:3}
\frac{\partial w^{(2)}(r_{12})}{\partial \mathbf{r}_{1}} 
-  \frac{\partial u(r_{12})}{\partial \mathbf{r}_{1}}
= \rho \int \frac{\partial u(r_{13})}{\partial \mathbf{r}_{1}} 
\frac{g^{(3)}(\mathbf{r}_{1},\mathbf{r}_{2},\mathbf{r}_{3})}{g(r_{12})} d\mathbf{r}_{3}\:,
\end{equation}
relating the difference between the mean force and the direct
pair-force to an integral over the force on particle 1 at
$\mathbf{r}_{1}$ due to a particle at $\mathbf{r}_{3}$, weighted by
the probability $\rho g^{(3)} d\mathbf{r}_{3}/g(r_{12})$ of finding a
particle in $d\mathbf{r}_{3}$ at $\mathbf{r}_{3}$ when it is known
that other particles are located at $\mathbf{r}_{1}$ and
$\mathbf{r}_{2}$.  This equation is a member of the Born-Green-Yvon
(BGY) hierarchy and is exact if pairwise interactions can be assumed.
Inserting $\beta w^{(2)}= - \ln g$ and using the KSA as a closure
relation, eq.~(\ref{eq:3}) yields the BGY integral equation for $g(r)$
\cite{egelbuch}.

\subsection{Kirkwood Superposition Approximation}
In a dilute system the overall interaction is dominated by the
interaction of individual pairs. Therefore the probability of finding
a certain arrangement of particles is nothing but the joint
probability of finding individual pairs. For a triplet with
particle distances $r$, $s$ and $t$ the approximate triplet
distribution function $g^{(3)}_{SA}(r,s,t)$ is then the product of the
pair distribution functions
\begin{equation}
g^{(3)}_{SA}(r,s,t)=g(r)g(s)g(t) \:.
\end{equation}
This is the so called Kirkwood superposition approximation (KSA)
\cite{kirkwood35}. Introducing the factor $G$ correcting the error
made by the KSA
\begin{equation}
\label{eq:1}
g^{(3)}(r,s,t)=g^{(3)}_{SA}(r,s,t)G(r,s,t)
\end{equation}
and the correction potential of mean force $\Delta w^{(3)}$
\begin{equation}
\label{eq:2}
\Delta  w^{(3)}(r,s,t) = - \ln G(r,s,t)/\beta
\end{equation}
we obtain for the triplet potential of mean force $w^{(3)}$
\begin{equation}
w^{(3)}(r,s,t) = w^{(2)}(r) + w^{(2)}(s) + w^{(2)}(t) + 
\Delta w^{(3)}(r,s,t) \:.
\end{equation}
All pair correlations in $g^{(3)}$ are included in $g^{(3)}_{SA}$
while $G$ quantifies the extent of intrinsic correlations due to the
simultaneous presence of a triplet of particles. For this reason, $G$
is called the triplet correlation function. In other words, if $G$ is
unity everywhere, there are no genuine triplet correlations in the
system, but only the trivial ones that can be expressed by pair
correlation functions. $\Delta w^{(3)}$ therefore measures the extra
correlation energy of three correlated particles relative to the
energy of superposed correlated pairs.  One should be careful not to
confuse $\Delta w^{(3)}$ with a real three-body potential $u^{(3)}$:
$\Delta w^{(3)}$ is a correlation energy and can thus have
non-vanishing values even in system with particles interacting
exclusively via pair-wise additive potentials $u(r)$, i.e., in cases
where higher-order potentials such as $u^{(3)}$ are strictly zero.

\section{Experimental systems and technical remarks}

\subsection{Two colloidal model systems}

The triplet functions that we here present are based on a large number
of particle configurations obtained from two different colloidal model
systems by means of video-microscopy. Both systems are two-dimensional
(or more precisely: quasi-2D). In the first system particles interact
via a tuneable magnetic dipole-dipole interaction and in the second
via an electrostatic double-layer interaction.

The first system consists of paramagnetic spherical colloidal
particles with an diameter of $4.7 \: \mu m$. They are located at the
bottom of an hanging water droplet whose surface can be accurately
controlled to be almost perfectly flat.  The positions of the
particles are recorded using digital video-microscopy with subsequent
image-processing on the computer. The field of view has a size of $520
\times 440 \: \mu m$ containing typically about $10^{3}$ particles. A
magnetic field $B$ applied perpendicular to the air/water interface
induces in each particle a magnetic moment $M= \chi B$ which leads to
a repulsive dipole-dipole pair-interaction energy of
\begin{equation}
\label{eq:ug}
\beta u(r) = \frac{\Gamma}{(\sqrt{\pi\rho}r)^{3}}
\end{equation}
with the interaction strength given by $\Gamma = \beta (\mu_{0}/4 \pi)
(\chi B)^{2} (\pi \rho)^{3/2}$. This is the only relevant contribution
to the interparticle-potential which is hence conveniently and
reversibly adjustable by varying $\Gamma$ through the external field
$B$. Note in particular that by introducing $\Gamma$ we have scaled
out the density. $\Gamma$ is thus the only parameter determining the
phase-behavior of the system: for $\Gamma < 57$ the system is liquid,
for $\Gamma > 60$ it is solid, and in between, i.e. for $57 < \Gamma <
60$, it shows an hexatic phase \cite{zahn99,zahn00}. The system can be
regarded as an almost ideal 2D model system as the out-of-plane motion
of the particles corresponds to less than 1 \% of their diameter.
Details about the preparation of the samples and the experimental
set-up can be found in \cite{zahn99,zahn00}.

The second system is an aqueous suspension of highly charged
sulphate-terminated PS particles of $\sigma = 3\: \mu m$ diameter,
confined between two glass plates with a 1 mm spacing. The particles
were furthermore exposed to vertical light forces which pushed them
toward the negatively-charged silica plate at the bottom of the cell,
confining the system more effectively to two dimensions. The particle
center positions were analyzed on-line with an imaging processing
software. The particle density was varied in the system
by a scanned optical laser tweezer which acts as a corral for the
investigated 2D colloidal suspension. The particles
interact with a screened Coulomb potential
\begin{equation}
\label{eq:uy}
\beta u(r)=\left(\frac{Z}{1+\kappa\sigma/2}\right)^2
e^{\kappa\sigma}\lambda_B \frac{e^{-\kappa r}}{r},
\end{equation}
with the (effective) particle charge $Z$, the particle diameter $\sigma$, the
Bjerrum-length $\lambda_B$ and the inverse Debye-screening-length
$\kappa$. Via the inversion of the Ornstein-Zernicke equation for a
low density system we can determine the potential quite accurately.
This experiment and also the inversion procedure has been described in
great detail in \cite{brunner,klein}. In fact, we here compute triplet
correlation functions based on exactly the same set of configurations
as already evaluated for other purposes in \cite{brunner,klein}.
Pair-correlation functions corresponding to the triplet correlation
functions of this work can also be found in \cite{brunner,klein}.

In both systems, we made sure that the system is well equilibrated and
used about 200 statistically independent configurations with
approximately 500-1000 particles.  We also performed standard
Monte-Carlo (MC) simulations, which offers us a third way to generate
the necessary set of particle configurations. The simulations produced
typically 500 configurations with 2000 particles yielding much better
statistics and the opportunity to check the experimental results for
statistical errors.

\subsection{How to compute triplet distribution functions}

The numerical procedure for computing $g^{(3)}$ is essentially the
same as the one used in simulations to determine $g(r)$.  For every
configuration we counted the triplets with sidelengths $r,s$ and $t$
in an array $[R_i,S_j,T_k]$ such that $R_i-\Delta R/2 \leq r <
R_i+\Delta R/2$
and accordingly for $s$ and $t$ with $\Delta R=\Delta S=\Delta T$. While normalizing
$g$ means each slot is divided by the area $A(R_i)=2\pi R_i \Delta R$ it
covers, we can write down in analogy an analytic expression for
$g^{(3)}$ valid for infinitesimal small $d\!R$:
\begin{equation}
\label{eq:normg3an}
A_{an}(R_i,S_j,T_k) = \frac{4\pi R_i S_j T_k d\!R^3}
{\sqrt{R_i^2 S_j^2 - 1/4 (R_i^2+S_j^2-T_k^2)^2}}\:.
\end{equation}
This expression also works fine for finite $\Delta R$ (replace $d\!R$
by $\Delta R$ in eq.~(\ref{eq:normg3an})) except in the case of
triangles with one angle close to 180 degrees. For example, a triangle
with $r=1.4 \Delta R, s = 1.4 \Delta R, t = 2.6 \Delta R$ would be
sorted into the slots $i=1, j=1, k=3$. The center values of these
slots obviously do not represent a triangle and thus the normalizing
procedure fails because the square root in eq.~(\ref{eq:normg3an})
turns imaginary. While we can neglect these slots for most of the
plots presented, we cannot ignore these counts when integrating over
the whole $g^{(3)}$ as it is done in the section on the Born-Green
equation. In these cases we are not allowed to replace $d\!R$ by
$\Delta R$ in eq.~(\ref{eq:normg3an}), but, instead, we have to
compute the normalization factor numerically. For a fixed distance $r$
one can easily calculate the intersection area $A(r,S_j,T_k)$ of two
circular rings with radii $S_j, T_k$ and width $\Delta R$. The numerical
normalization factor $A_{num}(R_i,S_j,T_k)$ is then
\begin{equation}
\label{eq:normg3num}
A_{num}(R_i,S_j,T_k) = 2\pi\int\limits_{R_i- \Delta R/2}^{R_i+\Delta R/2} r 
A(r,S_j,T_k) d\!r \:.
\end{equation}
More technical details can be found in the appendix of
\cite{krumhansl72a}.

\section{Results: Three-body distribution functions}

\begin{figure}
\includegraphics[width= \textwidth]{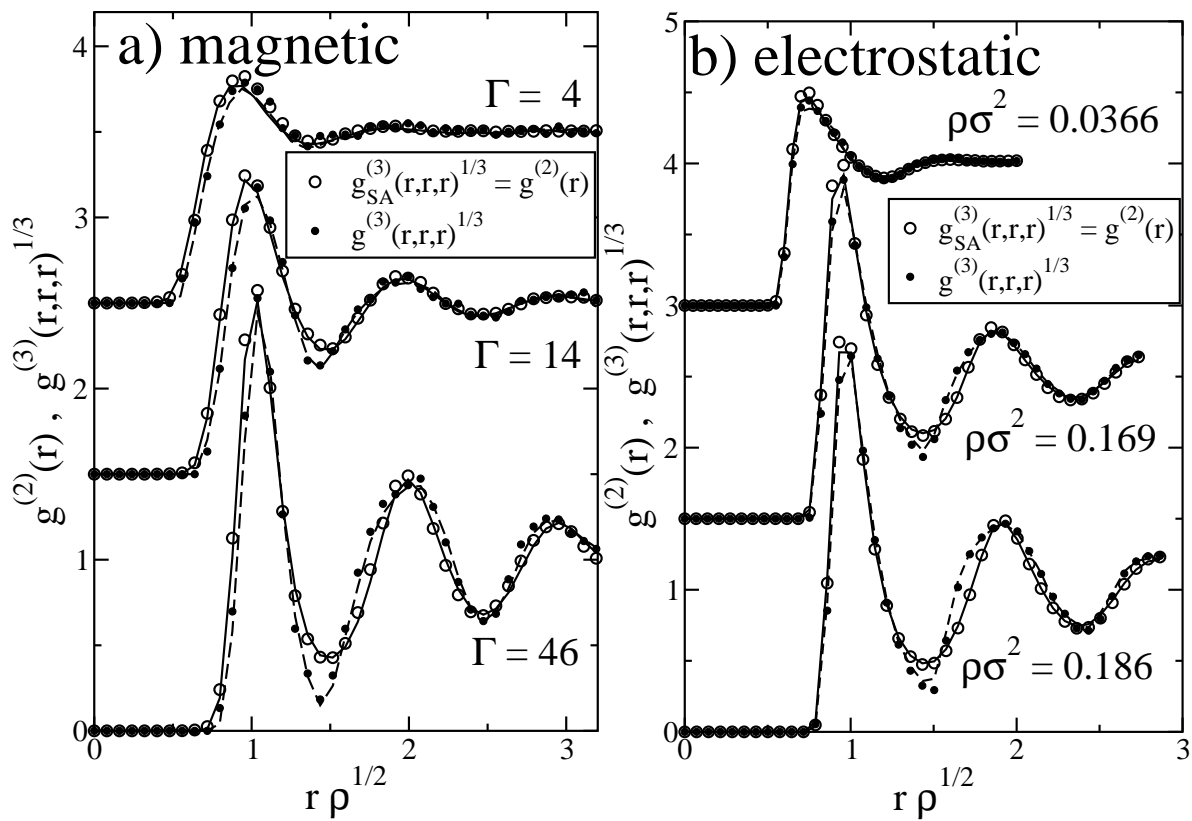}
\caption{\label{fig1} 
  Comparison between triplet distribution function
  $g^{(3)}(r,r,r)^{1/3}$ and the pair-distribution function $g(r)$ of
  (a) a paramagnetic colloidal fluid for different $\Gamma$, and (b)
  the charge-stabilized colloidal fluid for different colloid
  densities as indicated. Circles are experimental
  data, lines are simulation data. The curves for different densities
  (different $\Gamma$) are shifted for clarity.}
\end{figure}

We first concentrate on the equilateral triangle configuration. For
this special configuration we have $r = s = t$ and the superposition
approximation reduces to $g^{(3)}_{SA} = g(r)g(r)g(r)=(g(r))^3$.
Therefore, $(g^{(3)}_{SA}(r,r,r))^{1/3} = g(r)$ and a comparison of
$g(r)$ and $(g^{(3)}(r,r,r))^{1/3}$ reveals the importance of the
triplet correlation function $G(r,r,r)$ in eq.~(\ref{eq:1}).
Fig.~(\ref{fig1}) shows $(g^{(3)}_{SA}(r,r,r))^{1/3}$ and
$(g^{(3)}(r,r,r))^{1/3}$ for the system with the magnetic colloids
(Fig.~(\ref{fig1}.a)) and that with the charged colloids
(Fig.~(\ref{fig1}.b)). For each system type, we have analyzed three
different densities ($\Gamma$-values), where the systems are always
deep in the liquid phase. These are: $\Gamma = 4,14,46$ for the
magnetic system and $\rho\sigma^{2} = 0.037,0.167,0.186$ for the
charged system.

As expected, at low density (small $\Gamma$) Kirkwood's superposition
approximation holds very well. These are then weakly interacting
systems in which particles sparsely "meet" other particles and usually
interact only with a single other particle. The probability of finding
a certain triplet is thus nothing but the joint probability of finding
the appropriate pairs, and therefore the KSA is justified by the
independence of the particular pair probabilities.  Increasing the
interaction strength $\Gamma$ (i.e., the density in the charged
system) particles begin to interact with more than one other particle
at a time. The assumption that a third particle will not interfere
with the statistical distribution of the second particle begins to
deteriorate.  One can see that in Fig.~(\ref{fig1}): the first peak of
$g^{(3)}(r,r,r)^{1/3}$ is always reduced in height compared to $g(r)$
and the first minimum deviates even stronger. It is always more
pronounced for $g^{(3)}(r,r,r)^{1/3}$ and it changes its shape from
the sine-like oscillation of $g(r)$ to a more asymmetric one. This
seem to be just a packing effect which is relatively independent of
the interaction potential between the particles, as is evident from
the surprisingly close resemblance between the curves in
Fig.~(\ref{fig1}.a) and (\ref{fig1}.b). Note, however, that shape and
height of the first peak is different for both systems, as one would
expect for different pair-potentials. The dashed and solid lines
represent the results of our simulations, using the pair-potentials in
eq.~(\ref{eq:ug}) and (\ref{eq:uy}).  In the charged system, the
prefactor and the inverse screening length has been used as a
fit-parameter. The good agreement between simulation and experimental
data found for both the pair- and triplet distribution functions
demonstrates that the correlations can in both systems be understood
in terms of a picture of pair-wise interacting particles, although it
has to be emphasized that in the charged system the pair-potential is
an effective (density-dependent) one in which -- through the fitting
procedure -- many-body contributions are absorbed into the
pair-potential, see discussions in \cite{brunner,klein}. By contrast,
many-body interactions in the magnetic system can be safely ruled out
\cite{zahn99}.

\begin{figure}
\centering\includegraphics[width=0.5\textwidth]{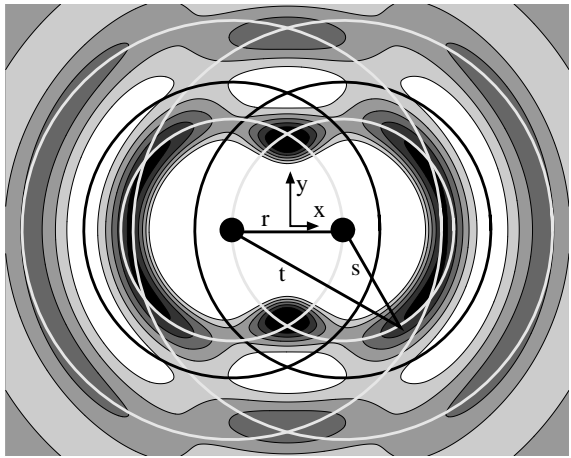}
\caption{\label{fig2} 
  2D-plot of the triplet distribution function in the Kirkwood
  superposition approximation, $(g^{(3)}_{SA}(r,s,t))^{1/3}$, for the
  $\Gamma=46$-measurement (magnetic system) with $r$ fixed at the most
  probable two-particle separation $r^{(2)}_{max}$. Darker areas
  indicate peaks in $g^{(3)}_{SA}$, circles represent the distance at which
  $g(r)$ around the respective particle has its first and second
  maximum (white circles) and its first minimum (black circles).}
\end{figure}
Another way to visualize $g^{(3)}(r,s,t)$ is to vary two variables
while fixing one variable which we here choose to be the variable $r$.
We take $r = r^{(2)}_{max}$ which is the distance where $g(r)$ has its
first peak (almost equal to $1/\sqrt{\rho}$). It is the most probable
distance between any pair of particles in the fluid. For this choice
the KSA leads to
\begin{equation}
g^{(3)}_{SA}(r^{(2)}_{max},s,t)= g(r^{(2)}_{max})g(s)g(t)
\label{eq:g3SArmax}
\end{equation}
and for $s\to\infty$ and $t\to\infty$ $g^{(3)}_{SA}$ goes to the
maximum value of pair correlation function,
$g^{(3)}_{SA}(r^{(2)}_{max},s,t)\to g(r^{(2)}_{max})$. To be
able to plot $g^{(3)}$ directly in the $(x,y)$-plane we have to
transform the variables $s$ and $t$ accordingly, $s=s(x,y)$ and
$t=t(x,y)$. Fig.~(\ref{fig2}) shows $g^{(3)}_{SA}(r^{(2)}_{max},s,t)$
for the $\Gamma=46$ measurement of the magnetic system in the $(x,y)$
plane (co-ordinate system as defined in the plot).  The lighter grey
to white areas correspond to values below and the darker grey to black
areas to values above the limiting value $g(r^{(2)}_{max})$.
Clearly, the plot is highly redundant, for symmetry reasons it would
have been sufficient to just show one quarter of it. In addition, we
show circles with radii corresponding to the first and second maximum
(white circles) and the first minimum (black circles) of $g(r)$ around
each particle.

With these circles the KSA prediction can be easily understood.  First
we expect a very pronounced peak where the two circles with the
shortest radii intersect. At this position is $s=r^{(2)}_{max}$ and
$t=r^{(2)}_{max}$; therefore both $g(s)$ and $g(t)$ are at their
maximum and that maximizes $g^{(3)}_{SA}(r^{(2)}_{max},s,t)$ in
eq.~(\ref{eq:g3SArmax}).  This position corresponds to an equilateral
triangle configuration. Accordingly, the first minimum on the $y$-axis
occurs where the two black circles intersect. Following the inner
white circle around the right particle means keeping $s=r^{(2)}_{max}$
constant. At the point where this line intersects the black circle --
indicating the first minimum of $g(t)$ -- we encounter a local minimum
along that line.  Following further the circle $s=r^{(2)}_{max}$ the
second maximum of $g(t)$ becomes important. But since the two circles
intersect almost tangential there is no distinctive structure but a
plateau-like ''banana'' enclosing the particle on the outside. In a
similar way, all other local maxima and minima in $g^{(3)}_{SA}$ can
be explained by combining the maxima and minima of $g(t)$ and $g(s)$.

\begin{figure}
\centering\includegraphics[width=\textwidth]{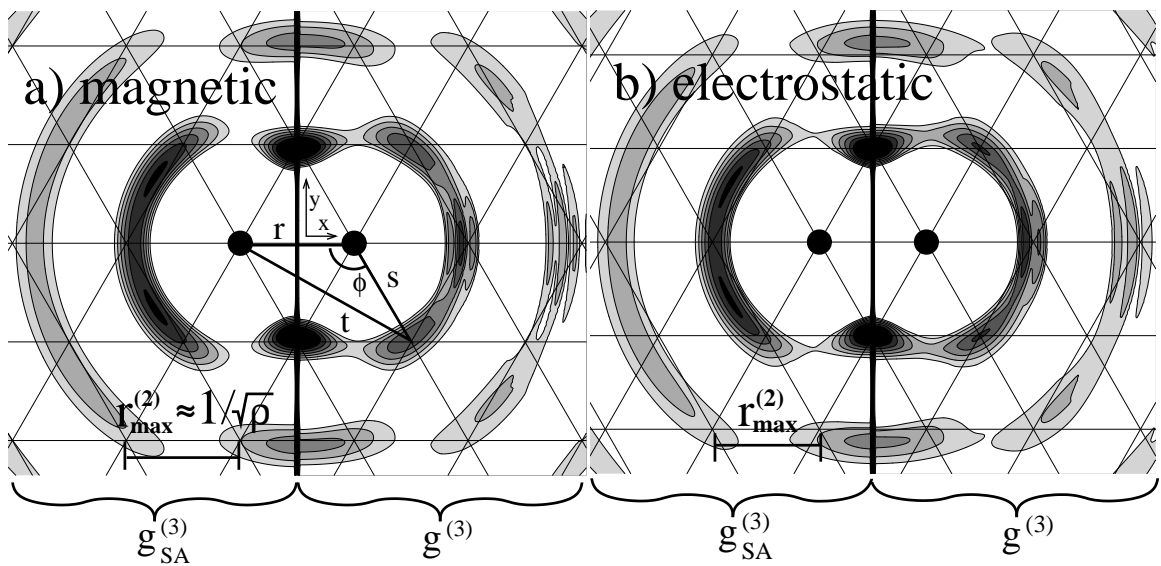}
\caption{\label{fig3}
Comparison between the full triplet distribution function, $g^{(3)}$,
  in the right half of each figure and the approximated triplet
  distribution function $g^{(3)}_{SA}$ (Kirkwood's superposition
  approximation) in the left half, for (a) a paramagnetic colloidal
  fluid at $\Gamma=46$, and (b) the charge-stabilized colloidal fluid
  at a density $\rho\sigma^{2} = 0.189$. The missing half of each
  distribution is just the mirror image of the one actually plotted.
  In contrast to Fig.~(\ref{fig2}), the constant $g(r^{(2)}_{max})$ is
  subtracted from the distributions and only positive values are
  plotted with a grey-level scheme between white (zero) and black
  (max. value). The underlying hexagonal lattice emphasizes approach to
  the crystalline phase.}
\end{figure}

Fig.~(\ref{fig3}) compares the full triplet correlation function
$g^{(3)}$ (right half of each figure) with the triplet correlation
function in the KSA, $g^{(3)}_{SA}$ (left half of each figure). Again,
we consider both the magnetic and the charged system, at the highest
density (highest $\Gamma$). To show the structure of $g^{(3)}$ more
clearly we have set the lower end of the grey-level code to the limiting
value $g(r^{(2)}_{max})$, so only the peaks, i.e. values where
$g^{(3)} > g(r^{(2)}_{max})$, are shown. The stripes that can be seen
especially close to the x-axis result from the transformation
$g^{(3)}(r_{max},s,t)$ to $g^{(3)}(r_{max},x,y)$ and appear due to
limited statistics. An hexagonal lattice with a lattice constant $a=
r^{(2)}_{max}$ is superposed. As in all our plots, we plot the cubic
root of $g^{(3)}$, to make the resulting numbers comparable to $g(r)$.

First of all one notices that the ''banana''-like structure is not
plateau-like anymore but shows a separation into three distinctive
peaks. For $\phi=120^\circ$ (the angle $\phi$ is defined in
Fig.~(\ref{fig3})) the peak matches almost exactly the grid point of
the underlying hexagonal lattice while for $\phi=180^\circ$ is
slightly shifted to the inside and is not so pronounced. In the shell of
second nearest neighbors there is also some inner structure
developing but a clear correspondence to the lattice points is not yet
recognizable. By contrast, there is no such correspondence between the
peak-structure of $g^{(3)}_{SA}$ and the lattice sites when the KSA is
applied. So this crystal-like local order that shows up already
between particles in the fluid phase, is obviously a genuine effect
of triplet correlations. Again, there is a remarkably close
resemblance between the results for the magnetic and the charged
system, demonstrating that these correlation effects have little to do
with the properties of the pair-potential.

\begin{figure}
\centering\includegraphics[width= \textwidth]{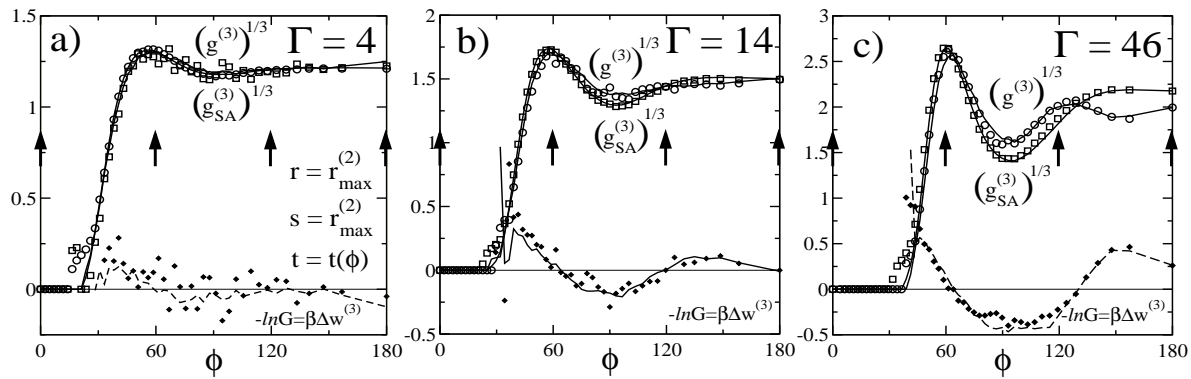}
\caption{\label{fig4} 
  $g^{(3)}$ and $g_{SA}^{(3)}$, for the magnetic system, from
  Fig.~(\ref{fig3}) for fixed values of $r=r^{(2)}_{max}$ and
  $s=r^{(2)}_{max}$, as a function of $t=t(\phi)$ with the angle
  $\phi$ as defined in Fig.~(\ref{fig3}).  Symbols (solid lines) for
  distributions generated from measured (MC-simulated) configurations.
  Also given is the logarithm of the triplet correlation function $G$,
  which is related to the triplet correlation energy $\Delta w^{(3)}$,
  see eq.~(\ref{eq:2}). Arrows indicate positions of lattice points of the
  hexagonal lattice in Fig.~(\ref{fig3}.)}
\end{figure}
To quantify our analysis, we now concentrate on the first neighbor
shell around the two central particles in Fig.~(\ref{fig3}), and fix
$r=r^{(2)}_{max}$ and $s=r^{(2)}_{max}$ while changing $t=t(\phi)$ by
varying the angle $\phi$ (defined in Fig.~(\ref{fig3})) between
$0^\circ$ and $180^\circ$. Fig.~(\ref{fig4}) shows
$g^{(3)}_{SA}(r^{(2)}_{max}, r^{(2)}_{max}, t(\phi))$ and
$g^{(3)}(r^{(2)}_{max}, r^{(2)}_{max}, t(\phi))$ as a function of
$\phi$, for the magnetic system. The corresponding results for the
charged system look again almost identical, and need not be further
considered. Varying $\phi$ and thus $t(\phi)$ while fixing
$r=r^{(2)}_{max}$ and $s=r^{(2)}_{max}$, we pass through all lattice
points of the particle's first coordination shell (arrows in
Fig.~(\ref{fig4}) mark positions of lattice points). It can be clearly
seen that at up to $\Gamma = 14$ $g^{(3)}$ and $g^{(3)}_{SA}$ are
almost identical and that at $\Gamma =46$ the full correlation
function $g^{(3)}$ shows a peak-structure in close correspondence with
the hexagonal structure. This is not the case for the the approximated
triplet distribution function $g^{(3)}_{SA}$.

Also given in Fig.~(\ref{fig4}) is the function $- \ln G$, i.e.
$\Delta w^{(3)}$, of eq.~(\ref{eq:2}). It is evident how $\Delta
w^{(3)}$ gradually forms on increasing $\Gamma$, with values up to one
$kT$ . It is also seen that the regions of attractive and repulsive
correlation energies $\Delta w^{(3)}$ correspond to the correcting
effect which the function $G$ has on $g_{SA}^{(3)}$ to ensure that
$g^{(3)}$ adapts locally to the hexagonal symmetry.  We conclude that
it is an effect entirely due to three-particle correlations, i.e. due
to the function $G$, which is responsible for the observed formation
of a crystal-like local environment around particles in a fluid well
below the freezing transition. All quantities displayed in
Fig.~(\ref{fig4}) are also compared to the corresponding results of
our MC simulations, showing again a remarkably good agreement.

\begin{figure}
\centering\includegraphics[width= 0.7\textwidth]{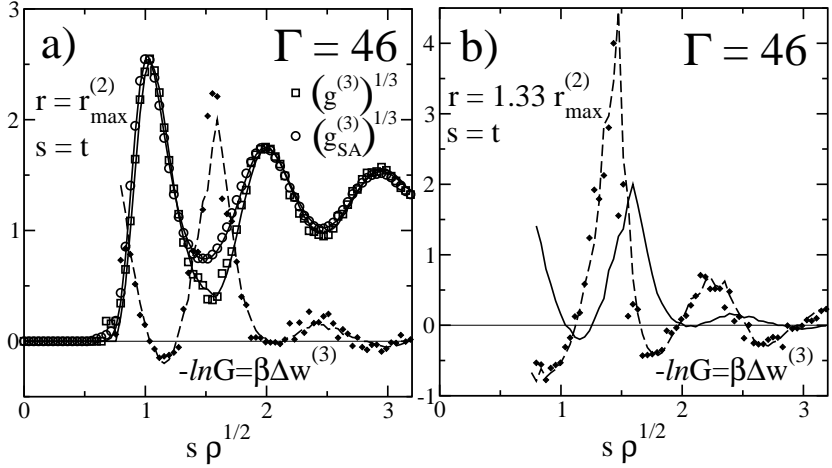}
\caption{\label{fig5} 
    (a) $g^{(3)}$ and $g_{SA}^{(3)}$, for the magnetic system, as in
  Fig.~(\ref{fig4}), but now for $r=r^{(2)}_{max}$ and $s=t$ (i.e.
  along the $y$-axis in Fig.~(\ref{fig3})). Symbols and lines as
  defined in Fig.~(\ref{fig4}).  (b) Solid line is the $\beta \Delta
  w^{(3)}$-curve of the left figure where $r=r^{(2)}_{max}$, compared
  here with the corresponding quantity for the case that $r = 1.33
  r^{(2)}_{max}$ (symbols: experiment, dashed line: simulation).}
\end{figure}
\begin{figure}
\caption{\label{fig6} 
Same plot as in Fig.~(\ref{fig3}) for the magnetic system ($\Gamma =
46$), but now for different distances $r$ between the two central
particles. Again, $g^{(3)}_{SA}$ in the left half of each figure, and 
$g^{(3)}$ in the right half. }
\end{figure}
While Fig.~(\ref{fig4}) follows $g^{(3)}$ along a circle around the
right particle in Fig.~(\ref{fig3}), Fig.~(\ref{fig5}) represents a
cut in the $y$-direction ($x$ = 0) in Fig.~(\ref{fig3}). In this
direction we have generally found the highest triplet correlation
energies $\Delta w^{(3)}$. For $r=r_{max}^{(2)}$ $\Delta w^{(3)}$ can
become as high as 2 $kT$, and if $r$ is slightly increased to $1.33
r_{max}^{(2)}$, we find an even higher correlation energy of more than
4 $kT$, see Fig.~(\ref{fig5}.b). Fig.~(\ref{fig6}) shows a sequence of
pictures similar to that in Fig.~(\ref{fig3}). While in
Fig.~(\ref{fig3}) we fixed $r$ to $r^{(2)}_{max}$, $r$ is now varied.
We expect triplet correlations to vanish if all three distances $r$,
$s$ and $t$ become large, and that then $g^{(3)} \to g^{(3)}_{SA}$.
Indeed, one observes from Fig.~(\ref{fig6}) that both halves of each
figure (displaying $g^{(3)}$ and $g^{(3)}_{SA}$) show more and more
resemblance the larger $s$ and $t$ are, i.e., in a region far away
from the central pair, and that the extent of this region increases
with increasing $r$. For example, for $r = 2 r_{max}^{(2)}$,
differences between both sides are to be seen only on the midplane.

\begin{figure}
\centering\includegraphics[width= 0.5\textwidth]{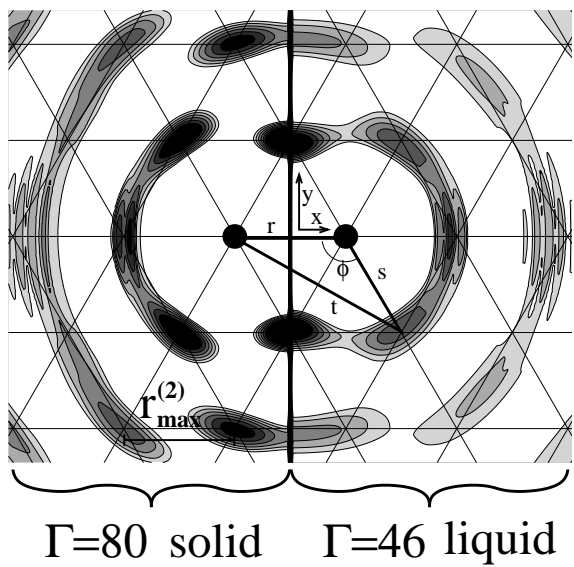}
\caption{\label{fig7} 
  Distribution functions $g^{(3)}(r=r^{(2)}_{max},s(x,y),t(x,y))$ for
  the magnetic system in the $(x,y)$-plane in the solid phase ($\Gamma
  = 80$, left half) and in the liquid phase ($\Gamma = 46$, right
  half), plotted in the same way as in Fig.~(\ref{fig3}). The $\Gamma
  = 80$ ($\Gamma = 46$) distribution is based on MC data (experimental
  data).  }
\end{figure}
The closer the system approaches the crystalline phase the more
pronounced do we expect the lattice points to be occupied. This is
tested in Fig.~(\ref{fig7}) which compares the triplet distribution
function for the system in the solid phase ($\Gamma = 80$) with that
for the system in the liquid phase ($\Gamma = 46$). For the solid
phase, we performed MC simulations, starting from a perfect hexagonal
lattice, while the $\Gamma = 46$ distribution is the same as in the
right half of Fig.~(\ref{fig3}.a). With regard to the correlations
between the central pair and the first coordination shell, there is
hardly any difference between the liquid and the solid phase.
Pronounced differences are observable, however, in the second shell:
in the liquid phase the next nearest neighbors are broadly distributed
midway between adjacent lattice nodes, while the $\Gamma = 80$
distribution clearly correlates much better with the lattice structure
(see, for example, the lattice point in the second shell next to the
mid-plane between the central pair).  However, even for $\Gamma = 80$
this correspondence is far from perfect; especially close to the $\phi
= 180^{o}$ direction there is still an extended smeared-out
distribution showing no clear preference for certain lattice points.
Clearly, approaching $T \to 0$ ($\Gamma \to \infty$), one will
ultimately observe peaks in $g^{(3)}$ positioned exclusively on the
lattice points.

\begin{figure}
\centering\includegraphics[width= 0.5\textwidth]{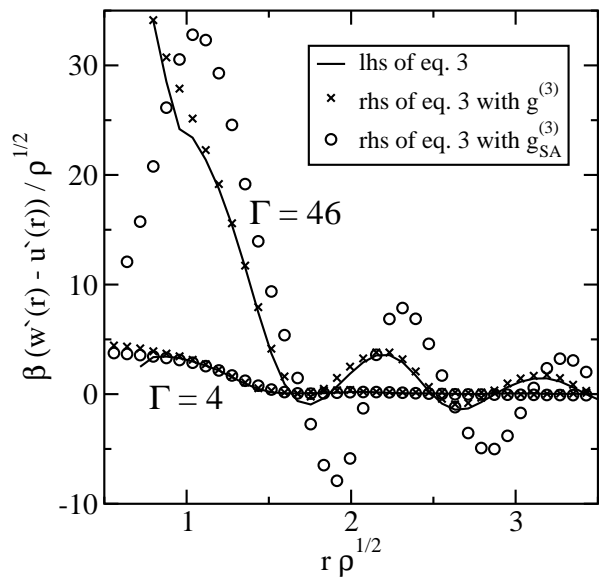}
\caption{\label{fig8} 
  Test of Kirkwood's superposition approximation (KSA) by means of the
  Born-Green equation, eq.~(\ref{eq:3}), using experimentally
  determined three-particle distribution functions ($\Gamma=46$ and
  $\Gamma=4$, magnetic system). Solid lines for the left hand side of
  eq.~\ref{eq:3}, symbols for the right hand side, evaluated using the
  full triplet distribution function $g^{(3)}$ (crosses) and the
  distribution function $g_{SA}^{(3)}$ in the KSA (open circles).}
\end{figure}
Krumhansl and Wang used eq.~(\ref{eq:3}) to check the accuracy of the
KSA in their early simulations of a LJ fluid
\cite{krumhansl72b,wang72}. We copy their idea, but use experimental
data to demonstrate the importance of three-particle correlations. We
numerically computed the right-hand side of eq.~(\ref{eq:3}) using
both the full and the approximated triplet function, $g^{(3)}$ and
$g_{SA}^{(3)}$, of the $\Gamma=4$ and $\Gamma=46$ measurement and
compare it in Fig.~(\ref{fig8}) to the left-hand side of
eq.~(\ref{eq:3}), evaluated using $u(r)$ and $g(r)$.  For the weakly
interacting system ($\Gamma=4$), $G$ is unity in value everywhere: 
triplet correlations are
unimportant, and, accordingly, the Born-Green equation can be
satisfied with triplet functions based on the KSA, see
Fig.~(\ref{fig8}). In the strongly interacting system ($\Gamma = 46$),
however, the KSA fails completely.  Three-particle correlations have
to be taken fully into account to obtain the correct difference
between mean and direct force via the Born-Green equation.  

A number of other approximations for $g^{(3)}$ are known
\cite{egelbuch}, and could be checked in a similar way. This includes
the Schofield equation \cite{schofield66}, relating $\partial g(r)/
\partial \rho$ to $g^{(3)}$, which in turn is the basic equation for a
number of thermodynamic consistency relationships \cite{egelbuch}.
From our results, we can expect that, here again, $g^{(3)}$ at high
$\Gamma$ cannot be approximated, but has to be taken in its full form.
Our results confirm predictions from simulation studies on LJ and
hard-sphere systems showing that qualitatively the structure of
$g^{(3)}$ is correctly described by the KSA, but that quantitative
failures can be appreciable
\cite{krumhansl72a,mcneil83,raveche74,medina98}.

\section{Conclusion and closing remarks}

We used experimental data obtained from digital video microscopy of
particles in a two-dimensional colloidal model fluid. From the
measured configurations, we subsequently calculated particle
correlation functions; in particular, triplet correlation functions,
containing much more information on the relative spatial arrangement
of particles than the pair-correlation function. We studied two
colloidal systems: one with particles interacting via a Yukawa
pair-potential, and another with particles interacting with a
$\Gamma/r^{3}$ potential. We have found very similar results for both
systems: When the liquid is near the freezing transition (high
density, high $\Gamma$), the deviations of the three-particle
correlation function from unity are considerable (with correlation
energies as high as 4~$kT$).  We observe the formation of a
crystal-like local environment around particles in a fluid well below
the freezing transition. Clearly, these are packing effects as is
evident from the close resemblance between the results for the two
systems having completely different pair-interaction potentials.  All
quantities examined have also been compared to results from MC
simulations, and showed always good agreement. We have finally used
our experimental data in combination with the Born-Green equation to
demonstrate that triplet correlations are also significant when
integrated over the whole volume. For the strongly interacting
magnetic system ($\Gamma = 46$), the KSA has been shown to fail
completely.  Three-particle correlations are thus seen to be important
not only to obtain locally the correct structure, but also to obtain
globally the correct difference between mean and direct force via the
Born-Green equation.

In principle, higher-order correlation functions could be computed
from the measured configurations in very much the same way, but it is
not clear to us what they could be useful for. We should also remark
that it soon will be possible to analyze higher-order correlation
functions also in 3D colloidal samples, with positional data recorded
using modern confocal microscope techniques. One has to be aware,
however, that in three dimensions a similar correspondence between the
peaks in $g^{(3)}$ and an underlying crystal lattice should be much
harder to find. In 3D, every triplet of particle lies, of course, also
in a plane, and can accordingly be plotted as in Fig.~(\ref{fig2}) and
(\ref{fig4}).  However, then there is not one, but a superposition of
many possible lattice planes that one has to compare this distribution
with. In this respect, the role of a crystal lattice in determining
the structure of a liquid is probably more pronounced in 2D than it is
in 3D.

\ack

HHvG and CR wish to thank Matthias Brunner and Prof. Clemens Bechinger
for all the many intensive discussions on this and other topics, for
their help and assistance, and the steady and close cooperation. We
gratefully acknowledge continuous support and stimulating discussion
with Prof. Rudolf Klein, and financial support from the Deutsche
Forschungsgesellschaft through the Konstanz SFB 513.

\end{document}